\documentclass{article}
\usepackage{spconf,amsmath,graphicx}
\usepackage{algorithm}
\usepackage[noend]{algpseudocode}
\usepackage{dirtytalk}
\usepackage{booktabs}
\usepackage[margin=1in]{geometry}
\usepackage{array}
\usepackage{amssymb}
\usepackage{url}
\newcolumntype{T}[1]{S[table-format=#1]}
\usepackage{lipsum}

\newcommand\blfootnote[1]{%
  \begingroup
  \renewcommand\thefootnote{}\footnote{#1}%
  \addtocounter{footnote}{-1}%
  \endgroup
}

\title{REAL-TIME HAND GESTURE RECOGNITION USING TEMPORAL MUSCLE ACTIVATION MAPS OF MULTI-CHANNEL SEMG SIGNALS}

\makeatletter
\def\@name{
\emph{Ashwin De Silva} $ ^1$\sthanks{these authors contributed equally to the work.}, 
\emph{Malsha V. Perera} $ ^{1 *}$, 
\emph{Kithmin Wickramasinghe} $ ^1$, 
\emph{Asma M. Naim} $^1$ \\
\emph{Thilina Dulantha Lalitharatne} $ ^2$, 
\emph{Simon L. Kappel} $^1$
}
\makeatother

\address{$^{1}$ Dept. Electronic and Telecommunication Eng., University of Moratuwa, Sri Lanka \\
        $^{2}$ Dept. Mechanical Eng., University of Moratuwa, Sri Lanka}

\begin{document}
%
\maketitle
\begin{abstract}
Accurate and real-time hand gesture recognition is essential for controlling advanced hand prostheses. Surface Electromyography (sEMG) signals obtained from the forearm are widely used for this purpose. Here, we introduce a novel hand gesture representation called Temporal Muscle Activation (TMA) maps which captures information about the activation patterns of muscles in the forearm. Based on these maps, we propose an algorithm that can recognize hand gestures in real-time using a Convolution Neural Network. The algorithm was tested on 8 healthy subjects with sEMG signals acquired from 8 electrodes placed along the circumference of the forearm. The average classification accuracy of the proposed method was 94\%, which is comparable to state-of-the-art methods. The average computation time of a prediction was 5.5ms, making the algorithm ideal for the real-time gesture recognition applications. 
\end{abstract}
\begin{keywords}
Multi-Channel Surface Electromyography, Real-Time Hand Gesture Recognition, Temporal Muscle Activation Maps, Onset Detection, Convolutional Neural Networks
\end{keywords}
\section{Introduction}
\label{sec:intro}

\blfootnote{© 2020 IEEE.  Personal use of this material is permitted. Permission from IEEE must be obtained for all other uses, in any current or future media, including reprinting/republishing this material for advertising or promotional purposes, creating new collective works, for resale or redistribution to servers or lists, or reuse of any copyrighted component of this work in other works.}

Surface Electromyogram (sEMG) signals from the forearm are widely used in gesture controlled applications and prosthesis control systems \cite{G2,J2}. The most commonly used gesture recognition method is based on extracting a set of temporal and frequency domain features from acquired sEMG recordings and then classifying them using different learning algorithms such as Support Vector Machines, Linear Discriminant Analysis (LDA) and Neural Networks \cite{F2,H2,I2,L2}.

A number of studies have been conducted to accurately classify hand gestures from the pre-recorded sEMG signals \cite{I2,B2,E2}. However, only a few studies \cite{A2,C2,D2} have focused on real-time hand gesture recognition using sEMG signals from the forearm. Most methods \cite{B2,A2,C2,K2} rely on binning of the sEMG signals, computing a set of features  (mean absolute value, waveform length, etc.) for each bin and labelling these bins using a trained classifier. Continuous binning causes real-time gesture predictions to be computationally expensive because it requires the algorithm to continuously predict the label of each bin despite the absence of a gesture onset. 

Most of the previous work on real-time hand gesture recognition have focused on extracting features by considering sEMG recordings from electrodes placed on the forearm as individual and uncorrelated entities. To the best of our knowledge,  only Furui et al. \cite{C2} have explored the correlations between the sEMG channels. However, Furui et al. performed gesture prediction for all bins, even when a gesture onset was not present.

In this paper, we propose the novel idea of TMA maps that can represent the individual and mutual activation patterns of forearm muscles. We then explore the potential of using TMA maps to detect the time of gesture onsets and to classify different gesture types. In addition, we introduce a novel TMA map based real-time hand gesture recognition algorithm which is computationally efficient and has an accuracy that is comparable to state-of-the-art real-time gesture classification algorithms.

\subsection{Method Overview}
\label{sec:format}

In traditional methods, features are usually extracted from individual channels, paying little or no attention to correlation between the channels. After careful observation of the signals, it was noted that multi-channel sEMG signals exhibit a correlated change across the channels, each time a hand gesture onset is occurring. Fig. 1(a) illustrates an sEMG recording from eight electrodes (channels) mounted along the circumference of the forearm. Looking at the recording, correlated signal changes are observed at the time of the onsets A and B, corresponding to middle finger flexion and middle finger extension. Fig. 1(b) illustrates the envelopes of two sEMG recordings. By extracting the envelope of an sEMG signal, a measure of the amplitude modulation of the signal is obtained. The signal amplitude represents the activity/tension of the muscle mass. Therefore, correlated changes in envelopes of the sEMG signals, across a given set of channels, represent the pattern of activated muscles related to a hand gesture. In Fig. 1(b), it can be observed that these correlated changes are unique to different gestures.

\setlength{\textfloatsep}{15pt}

\begin{figure}[!t]
\begin{minipage}[b]{1.0\linewidth}
  \centering
  \centerline{\includegraphics[width=\columnwidth]{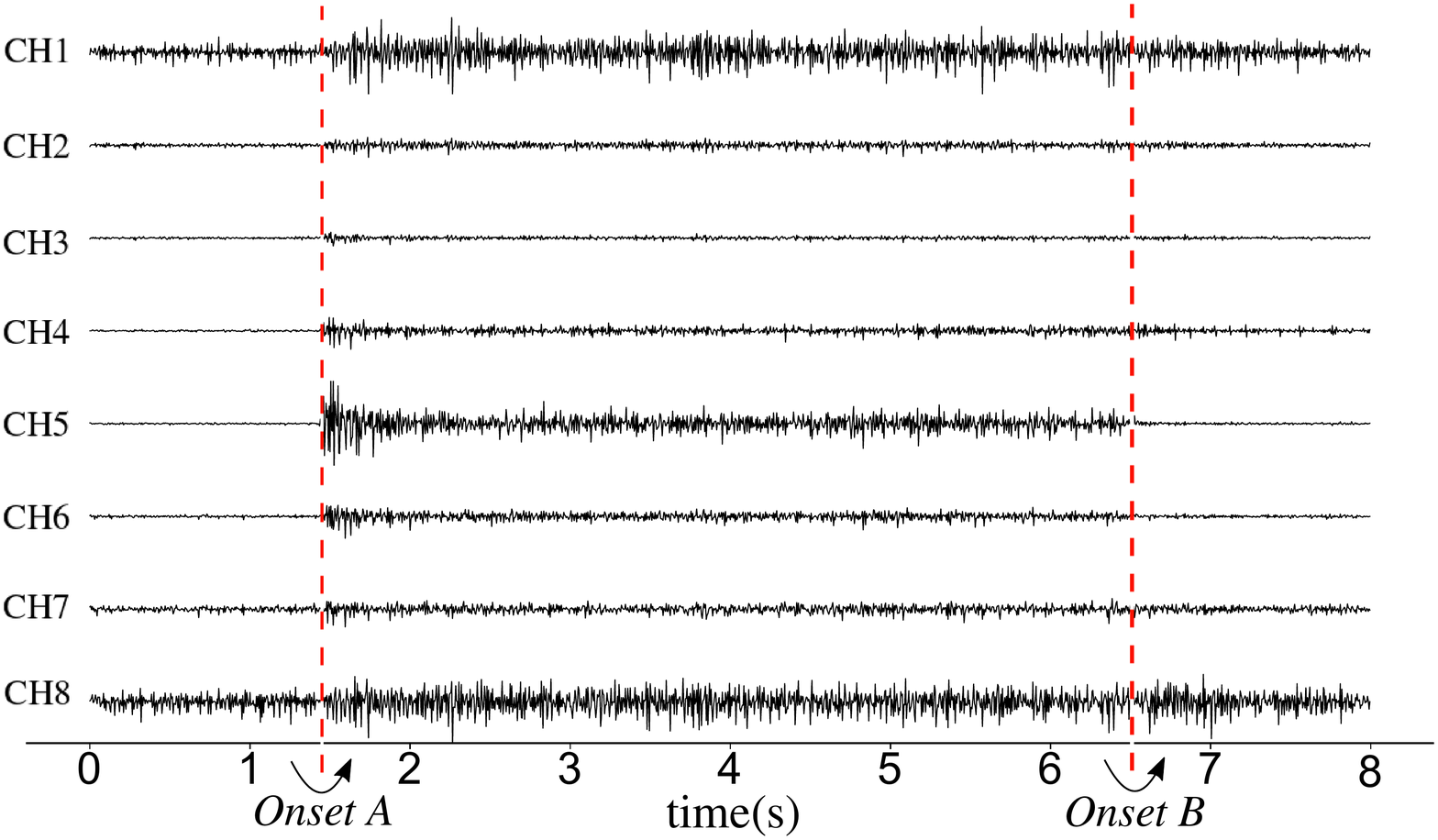}}
  \centerline{(a)}\medskip
\end{minipage}
\begin{minipage}[b]{1.0\linewidth}
  \centering
  \centerline{\includegraphics[width=\columnwidth]{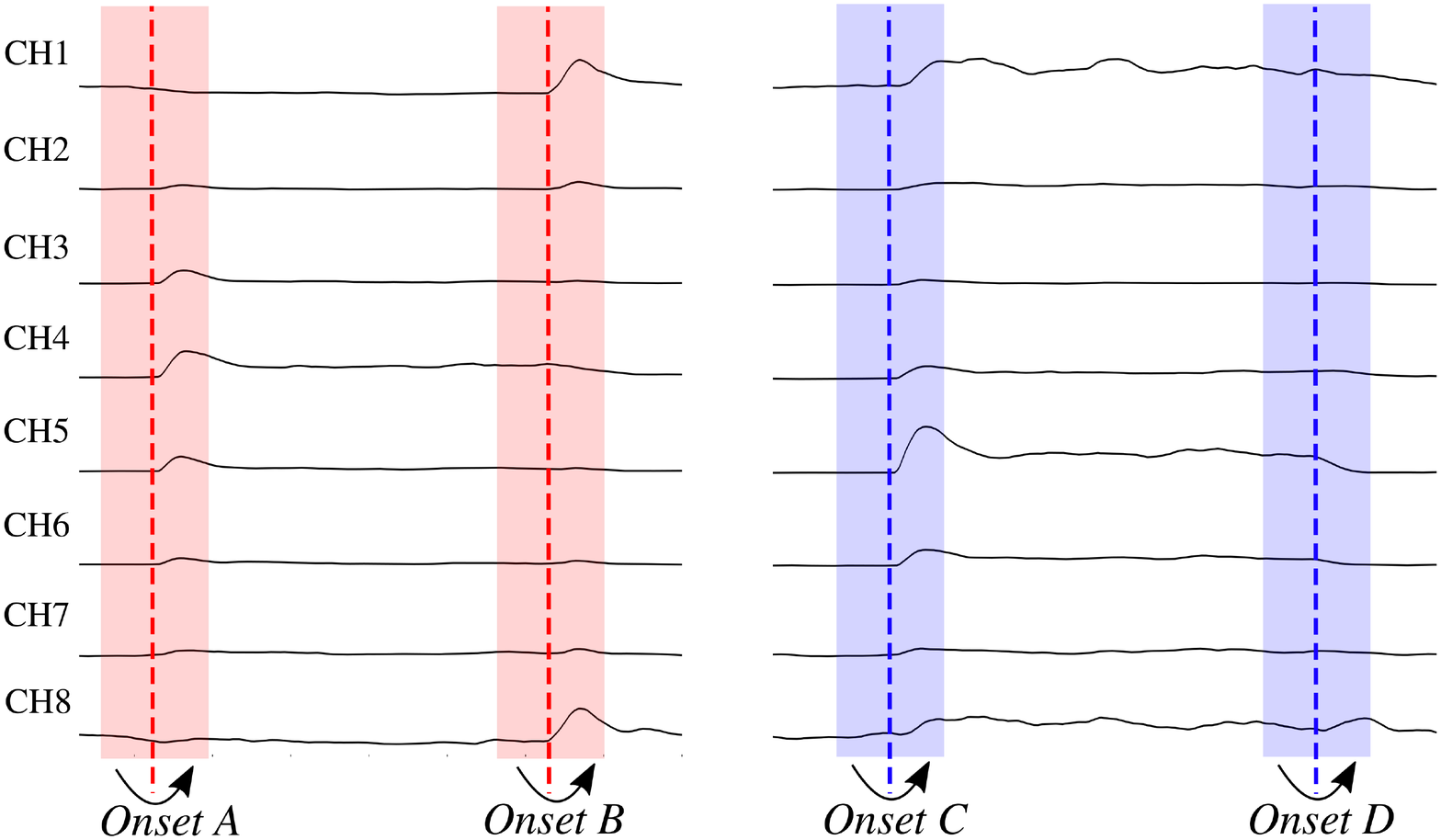}}
  \centerline{(b)}\medskip
\end{minipage}
\vspace{-2em}
\caption{(a) Raw sEMG signals. Onsets A and B correspond to middle finger flexion and extension, respectively. (b) The envelopes of two sEMG recordings. Onsets A, B, C, and D correspond to middle finger flexion, middle finger extension, ring finger flexion and ring finger extension, respectively.}
\label{fig:res}
\end{figure}

The correlated changes were described in a TMA map, which quantify the temporal changes in individual and mutual muscle activations within a time window of fixed size. Thus, a time series of TMA maps is an estimate of the muscle activations over time. It was expected that significant changes would be present in the TMA map at the occurrence of a gesture onset, and that this change could be quantified to determine the time of the gesture onset. Also, it was hypothesized that each hand gesture had a unique TMA map which could be used to classify the gesture.

\section{METHOD}
\label{sec:pagestyle}

\subsection{Generating Temporal Muscle Activation Maps}

Initially, the envelopes of the signals were extracted by performing full-wave rectification and low pass filtering at $f_c$ Hz using a second order Butterworth filter.

Let $x_l(n)$, $n = 0,...,N-1$, be $N$ samples of the envelope of an sEMG signal recorded using electrode $l$ of $L$ electrodes. The column vector $\textbf{a}(n)$ is given by
\begin{multline}
\begin{aligned}
\textbf{a}(n) = \big [ &x_0(n), x_1(n), \dots , x_{L-1}(n),\\
        &x_0^2(n), x_0(n)x_1(n), \dots , x_0(n)x_{L-1}(n),\\
		&x_1^2(n) , x_1(n)x_2(n) ,\dots , x_1(n)x_{L-1}(n),\\
		\dots,&x_{L-2}^2(n), x_{L-2}(n)x_{L-1}(n), x_{L-1}^2(n) \big ]^T
\end{aligned}
\end{multline}
Then, the TMA map is defined as the matrix, $\textbf{A}(n)$, formed by the column vectors $\textbf{a}(n)$ to $\textbf{a}(n+M-1)$
\begin{equation}
\textbf{A}(n) = [\textbf{a}(n) \text{   }\textbf{a}(n+1) \text{   }  \dots \text{   } \textbf{a}(n+M-1)]
\end{equation}
where, M is the window length in samples. The sketch in Fig. 2 illustrates the process of a generating TMA map.

\begin{figure}[!b]

\begin{minipage}[b]{1.0\linewidth}
  \centering
  \centerline{\includegraphics[width=7.5cm]{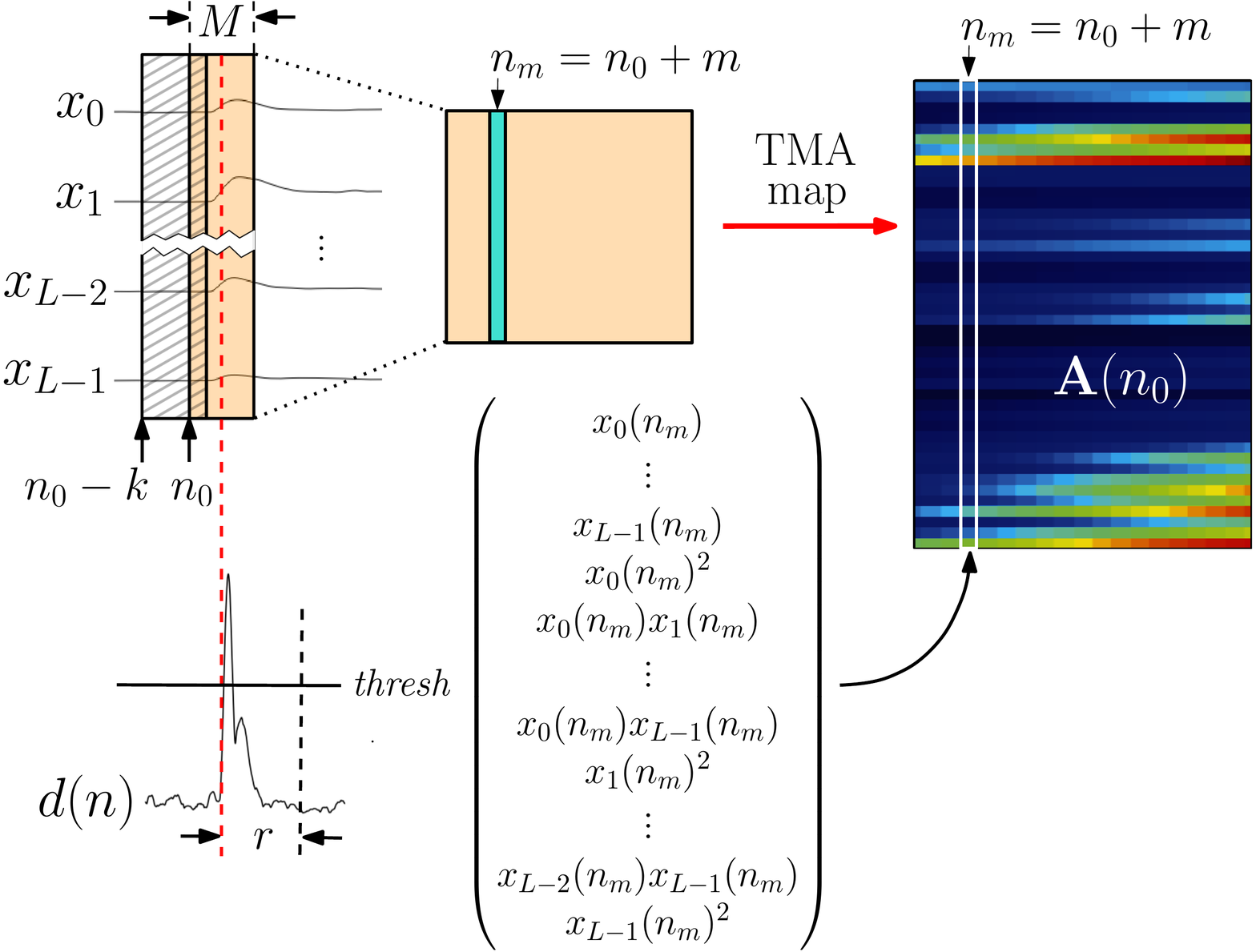}}
\end{minipage}
\caption{Process of generating TMA maps and detecting onsets }
\label{fig:res}
\end{figure}

\subsection{Onset Detection}

Earlier, we assumed that the muscle activation patterns significantly change at a gesture onset. To quantify the temporal change of muscle activation patterns, we define a difference signal $d(n)$ which is based on the TMA maps that was introduced in the previous section.
\begin{equation}
d(n) = ||\textbf{A}(n) - \textbf{A}(n-k)||_F
\end{equation}
where $||.||_F$ is the Frobenius norm \cite{M2} of a matrix and $k$  is the number of samples between two adjacent TMA maps. 

As indicated from its definition, it can be expected that $d(n)$ increases when a change in the muscle activations occur, corresponding to a gesture onset.  During experimentation, this was confirmed with prominent peaks at the times of gesture onsets. A sketch of $d(n)$ is given in Fig. 2. The gesture onsets were detected using a threshold that was estimated using the $d(n)$ values computed from the training data. After a gesture was detected, the onset detection procedure was paused for a duration of $r$ samples. We refer to this pause as the refractory period, because it was introduced to minimize false-positive onset detection. 

\subsection{Classification }

\subsubsection{Training the Classifiers}

As mentioned in section 1.1, the TMA maps at the time of gesture onsets can be used as training data for a classification algorithm. However, it is likely that the TMA maps in the neighborhood of the detected onset would also represent the muscle activity associated with the corresponding hand gesture. Thus, the TMA maps for training of the classifier were extracted in a window around the detected onset time. The window was centered at the sample time of the detected onset, and had a width of $\delta$ samples. To maximize the amount of training data, TMA maps were extracted for all sample times in the window. These TMA maps together with their corresponding hand gesture labels, formed the dataset for training of the classifier.

Because of the image-like nature of the TMA maps, we considered it to be beneficial to use a Convolutional Neural Network (CNN) to classify the hand gestures. The network architecture included two convolutional layers followed by two fully connected layers. The convolutional layers in a CNN serve as feature extractors, and thus they learn the feature representations of their input images \cite{N2}. Because TMA maps are images that embed the temporal variation of both individual and mutual muscle activation patterns, the convolutional layers of the CNN can learn abstract features from the maps without having to go deeper than two layers. The fully connected layers that follow the convolutional layers interpret these features and perform high-level reasoning that can associate the extracted features with the corresponding hand gestures. 

Prior to training, the two regions corresponding to the first and second order terms of the TMA maps were normalized separately to the interval $[0, 1]$.

\subsubsection{Real-Time Hand Gesture Prediction}

The trained classifier was used for real-time gesture prediction. Gesture prediction was only performed when an onset was detected. The prediction algorithm is summarized in Algorithm 1.

\begin{algorithm}[!t]
\caption{Real-Time Hand Gesture Recognition}
\label{euclid}
\begin{algorithmic}[1]
\State $n = 0, \text{ } \textbf{A}(0) \gets \textit{initial TMA map}$
\State \emph{top}:
\State $\text{wait for k new samples of the sEMG signal envelopes}$
\State $n \gets n+k$
\State $\textbf{A}(n) \gets \textit{TMA map at } n$
\State $d(n) \gets ||\textbf{A}(n) - \textbf{A}(n-k)||_F $
\If {$(d(n) > \textit{thresh}) \land (e.t.p  > r)$}
\State $\textit{prediction}  \gets \textit{CNN } \text{(} \textbf{A}(n) \text{)} $
\EndIf
\State \textbf{goto} \emph{top}.
\end{algorithmic}
$e.t.p$ denotes the elapsed time since last prediction
\end{algorithm}

\section{Experiment}

The objective of the experiment was to assess the accuracy of the proposed gesture onset detection and gesture classification method. The 5 gestures illustrated in Fig. 3 were included in the experiment; middle finger flexion, ring finger flexion, v-flexion, hand closure and pointer. Raw sEMG signals were recorded at a sampling rate of 200 Hz from 8 healthy subjects (4 males and 4 females, age : $25 \pm 2$) using the  commercially available Myo Armband (Thalamic Labs, Canada). The placement of the Myo Armband on the forearm is shown in the Fig. 3. 

\begin{figure}[!htb]

\begin{minipage}[b]{1.0\linewidth}
  \centering
  \centerline{\includegraphics[width=6.9cm]{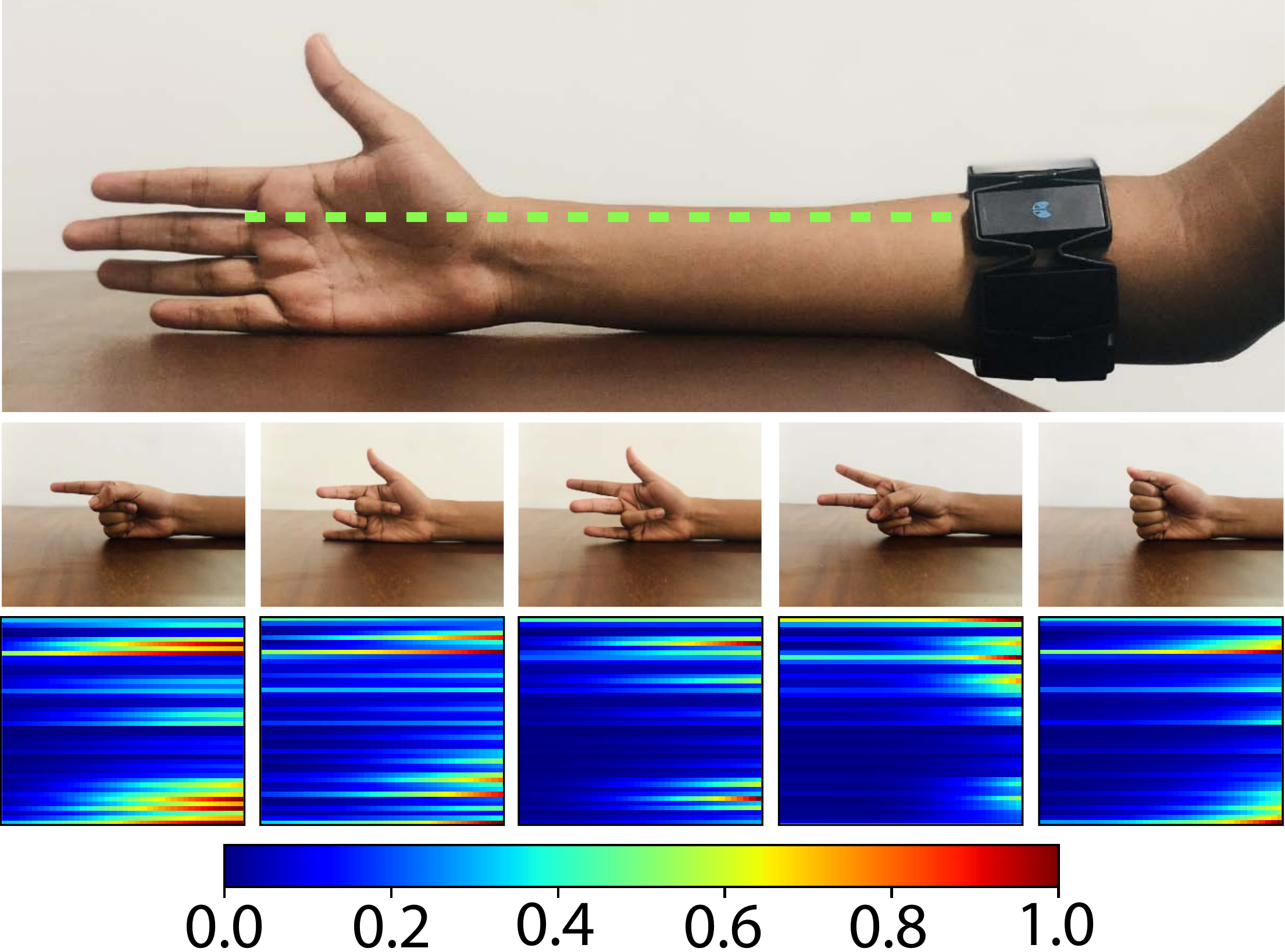}}
\end{minipage}
\vspace{-1.6em}
\caption{Top row : The hand in the neutral position. The Myo Armband was placed so that the base of the middle finger was in line with the Myo Armband logo, as indicated by the green dashed line. Middle row : Types of hand gestures. In order from left to right; pointer, middle finger flexion, ring finger flexion, v-flexion, hand closure. Bottom row : The TMA maps at the onsets of the corresponding gestures.}
\label{fig:res}
\end{figure}

During the data collection, the subjects were asked to perform 20 repetitions of each gesture with their right hand. Each gesture was held for a period of 5 seconds and was followed by a resting period of 5 seconds. During the resting period, the subjects were asked to keep their hand in a neutral position with relaxed hand muscles. The experiment was approved by the Ethics Review Committee at University of Moratuwa (Ethics Review Number : ERN/2019/007).

The following parameters were used to compute the TMA maps ; $f_c = 2$ Hz, $M = 80$ samples ($0.4$ s) and $k  = 20$ samples ($0.1$ s). When 5 recordings of each gesture type were acquired, the variance, $\sigma_d^2(g)$, of $d(n)$ was computed separately for each gesture type. These variances were used to determine the onset detection threshold, which was given by $4 \times \frac{1}{G} \sum_{g=1}^{G} \sigma_d(g)$ where $G$ is the number of gesture types. The refractory period was set to 2 s ($r=400$ samples). When forming the training datasets, a window width of $\delta = 120$ samples ($0.6$ s) was used.

It should be noted that the extension gestures were not considered for this study. It was assumed that all the flexion gestures were followed by corresponding extension gestures, bringing the hand back to the neutral position. Therefore, the onsets immediately after a flexion gesture were exempted from classification.

After extracting a training dataset of TMA maps for each of the subjects, individualized CNNs (implemented using Keras) were trained with the Stochastic Gradient Descent optimizer using the following parameters; learning rate of 0.001 for 15 epochs, filter size of $3 \times 3$ for both convolutional layers, 100 and 20 neurons in the first and second fully connected layers, respectively. The CNN models were evaluated in a real-time study containing a random sequence of 150 hand gestures, ensuring that each gesture type would have an equal number of repetitions.

\vspace{-1em}
\section{RESULTS}
\label{sec:prior}

Table 1 reports the classification accuracies obtained from the proposed real-time gesture recognition algorithm and compare them with state-of-the-art methods described by Crepin et al. \cite{A2} and Furui et al. \cite{C2}.

Crepin et al. used a continuous binning approach and classified each bin with a LDA classifier considering channels as uncorrelated entities. Furui et al. used a muscle synergy based approach where the classification was performed for each bin by a recurrent log-linearized Gaussian mixture network. Our method was able to recognize the gestures middle flexion and ring flexion with higher accuracies than these state-of-the-art methods. Compared to Crepin et al., the classification accuracy of our method was higher for all gestures. This could be because we considered both individual and mutual muscle activations. When comparing to Furui et al., it is important to note that the Myo Armband has the inherent limitation that all the electrodes are placed on a common circumference around the forearm. This limits the access to some of the muscles that are directly linked to certain finger motions. Even with this limitation, we obtained results that are comparable to the results by Furui et al., who used optimal electrode positions on the forearm. We speculate that the number of recognizable gestures and the gesture classification accuracy could be significantly improved if optimal electrode positions on the forearm were used.
\vspace{-1em}
\begin{table}[!h]
\setlength\tabcolsep{0pt}
\caption{Classification Accuracy (\%)} \label{tab:freq}
\centering
\smallskip
\begin{tabular*}{\columnwidth}{@{\extracolsep{\fill}}l cccc}
\toprule
Hand & {Crepin}& {Furui} & {Proposed} \\
Gestures & {et al. \cite{A2}}& {et al. \cite{C2}} & {Method} \\
\midrule
  Middle Flexion & 81.90   & 91.30    & \textbf{96.67}     \\
  Ring Flexion & 93.50   &   -   & \textbf{94.58}     \\
  Hand Closure & 77.87   & \textbf{97.01}     & 93.75    \\
  V-Flexion& -   & \textbf{95.06}     & 92.91     \\
  Pointer& 80.85   & \textbf{97.27}     & 92.50    \\
  \midrule
  Total & 83.53 & \textbf{95.16}   & 94.08    \\
\bottomrule
\end{tabular*}
\end{table}

\begin{figure}[!h]
\vspace{-1em}
\begin{minipage}[htb]{1.0\linewidth}
  \centering
  \centerline{\includegraphics[width=7.5cm]{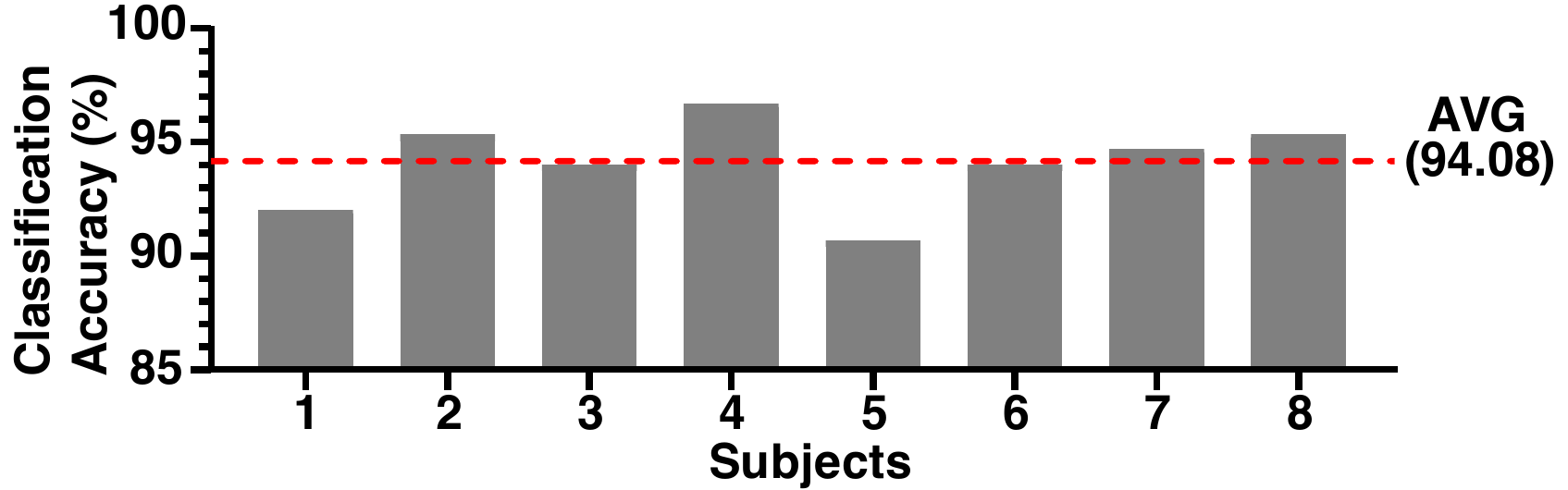}}
\end{minipage}
\vspace{-1em}
\caption{Classification accuracies of the proposed method across the subjects.}
\label{fig:res}
\end{figure}

In Fig.4, we report the subject-wise classification accuracy for the proposed method. Note that the classification accuracy was above 90\% for all subjects.  

The average computation time of a gesture prediction was 5.5 ms for the proposed method executed on a computer equipped with a 2-core 2.4 GHz Intel Core i7 CPU. Given that the computation time was significantly lower than the time between TMA map extraction, $k$, the algorithm supported real-time execution.

\vspace{-0.5em}

\section{CONCLUSION}

In this paper, we have introduced Temporal Muscle Activation (TMA) maps that represent the activation patterns of forearm muscles. The TMA maps were utilized to detect gesture onsets and recognize five selected hand gestures in real-time with an accuracy that was comparable to the state-of-the-art methods. Looking forward, it might be possible to increase the classification accuracy even further with more optimally placed electrodes.

\section{ACKNOWLEDGEMENTS}
\label{sec:refs}

Authors extend their gratitude to the Bionics Laboratory of Dept. of Mechanical Eng. at University of Moratuwa. 


\bibliographystyle{IEEEbib}
\bibliography{refs}

\end{document}